%% file: main.tex
\newif\ifarxiv
\arxivtrue 
\ifarxiv
    \documentclass{article}
\else
    \documentclass{elsarticle}
\fi

\usepackage[english]{babel}
\usepackage{amssymb,amsmath,amsthm}
\usepackage{comment}
\usepackage{microtype}
\usepackage[numbers]{natbib}
\usepackage{graphicx}
\usepackage[colorinlistoftodos]{todonotes}
\usepackage{url}
\usepackage{threeparttable}

\newcommand{\NP}{\ensuremath{\mathcal{NP}}}

\newcommand{\ver}[2]{\ensuremath{#1_{\vert #2}}}
\newcommand{\CEVS}{\textsc{Cluster Editing with Vertex Splitting}}

\newcommand{\titl}{\title{Cluster Editing with Vertex Splitting
}} 

\let\svthefootnote\thefootnote
\newcommand\blankfootnote[1]{%
  \let\thefootnote\relax\footnotetext{#1}%
  \let\thefootnote\svthefootnote%
}

\usepackage{subcaption}

\usepackage{tikz}
\usetikzlibrary{graphs,graphs.standard,calc,positioning}
\tikzset{%
    added/.style={green!60!black, densely dashed, thick},
    deleted/.style={red!70!black, loosely dotted, very thick}
}

\newtheorem{rul}{Reduction Rule}

    {\end{tabular}\end{center}\end{sspacing}}

\newtheorem{lemma}{Lemma}
\newtheorem{corollary}{Corollary}
\newtheorem{theorem}{Theorem}
\newtheorem{definition}{Definition}

\newcommand{\poly}{\text{poly}}

\usepackage[hidelinks]{hyperref}
\usepackage[capitalize,noabbrev]{cleveref}
\usepackage{bookmark}

\ifarxiv
    \usepackage[a4paper,top=3cm,bottom=3cm,left=3cm,right=3cm,marginparwidth=2cm]{geometry}
    \newtheorem{open}{Open Problem}
\else
    \newdefinition{open}{Open Problem}
    \newdefinition{remark}{Remark}
    \newproof{proof}{Proof}
    
    \author[LAU]{Faisal N. Abu-Khzam}
    \ead{faisal.abukhzam@lau.edu.lb}
    \author[Trier]{Emmanuel Arrighi}
    \ead{emmanuel@arrighi.eu}
    \author[Bergen]{Matthias Bentert}
    \ead{matthias.bentert@uib.no}
    \author[Bergen]{Pål Grønås Drange}
    \ead{Pal.Drange@uib.no}
    \author[CDU]{Judith Egan}
    \ead{judith.egan@cdu.edu.au}
    \author[UNSW]{Serge Gaspers}
    \ead{serge.gaspers@unsw.edu.au}
    \author[UNSW]{Alexis Shaw}
    \ead{Alexis.Shaw@student.unsw.edu.au}
    \author[OJL]{Peter Shaw}
    \ead{petershaw@ojlab.ac.cn}
    \author[Utah]{Blair D. Sullivan}
    \ead{sullivan@cs.utah.edu}
    \author[Bergen]{Petra Wolf}
    \ead{mail@wolfp.net}
    
    \address[LAU]{Lebanese American University, Lebanon}
    \address[Bergen]{University of Bergen, Norway}
    \address[Trier]{University of Trier, Germany}
    \address[CDU]{Charles Darwin University, Australia}
    \address[UNSW]{The University of New South Wales, Australia}
    \address[OJL]{Oujiang Laboratory, Wenzhou, Zhejiang, China}
    \address[Utah]{University of Utah, USA}

\fi

\begin{document}

\ifarxiv
    \date{}
    \titl
    \author{Faisal N. Abu-Khzam \and Emmanuel Arrighi \and Matthias Bentert \and Pål Grønås Drange \and Judith Egan \and Serge Gaspers \and Alexis Shaw \and Peter Shaw \and Blair D. Sullivan \and Petra Wolf}

    \maketitle
\else
    \begin{frontmatter}
    \titl
\fi


\begin{abstract} 
\textsc{Cluster Editing}, also known as \textsc{Correlation Clustering}, is a well-studied graph modification problem. 
In this problem, one is given a graph and the task is to perform up to $k$ edge additions or deletions to transform it into a cluster graph, i.e., a graph consisting of a disjoint union of cliques.
However, in real-world networks, clusters are often overlapping.
For example in social networks, a person might belong to several communities---e.g.\ those corresponding to work, school, or neighborhood.
Other strong motivations come from biological network analysis and from language networks. Trying to cluster words with similar usage in the latter can be confounded by homonyms, that is, words with multiple meanings like ``bat''.
In this paper, we introduce a new variant of \textsc{Cluster Editing} whereby a vertex can be split into two or more vertices.
First used in the context of graph drawing, this operation allows a vertex~$v$ to be replaced by two vertices whose combined neighborhood is the neighborhood of~$v$ (and thus $v$ can belong to more than one cluster).
We call the new problem \textsc{Cluster Editing with Vertex Splitting} and we initiate the study of it.
We show that it is \NP-complete and fixed-parameter tractable when parameterized by the total number~$k$ of allowed vertex-splitting and edge-editing operations.
In particular, we obtain an~$O(2^{9k \log k} + n + m)$-time algorithm and a $6k$-vertex kernel.
\end{abstract}

\ifarxiv
    \blankfootnote{Author E-Mail addresses: faisal.abukhzam@lau.edu.lb, emmanuel.arrighi@gmail.com,
    matthias.bentert@uib.no,
    pal.drange@uib.no,
    judith.egan@cdu.edu.au,
    serge.gaspers@unsw.edu.au,
    Alexis.Shaw@student.unsw.edu.au,
    petershaw@ojlab.ac.cn,
    sullivan@cs.utah.edu, and~mail@wolfp.net.}
\else
    \end{frontmatter}
    \clearpage
\fi



\section{Introduction}
\textsc{Cluster Editing} is defined as follows.
Given a graph $G$ and a non-negative integer $k$, one is asked whether~$G$ can be turned into a disjoint union of cliques by a sequence of at most $k$ edge-editing operations (i.e. additions or removals of edges).
The problem is known to be \NP-complete since the work of K\v{r}iv\'{a}nek and Mor\'{a}vek~\cite{KM86}
and it is fixed-parameter tractable when parameterized by $k$, the total number of allowed edge-editing operations \cite{Bocker12,Cai96,Gramm05}.
Over the last decade, \textsc{Cluster Editing} has been well studied from both theoretical and practical perspectives~\cite{Bocker12,ChenMeng,Chen2006,DAddario2014,dehne2006cluster, fadiel2006computational,fellows2007efficient,guo2009more,Huffner2010}. 

In general, clustering results in a partitioning of the input graph.
Hence, it forces each vertex to be in one and only one cluster.
This can be a limitation when the entity represented by a vertex plays a role in multiple clusters.
This situation is recorded in work on gene regulatory networks \cite{abu2005}, where enumeration of maximal cliques was considered a viable alternative to clustering.
Moreover, such vertices can effectively hide clique-like structures and also greatly increase the computational time required to obtain an optimal solution~\cite{radovanovic2010hubs,tomavsev2011role}.

In this paper, we introduce a new variant, which we call \textsc{Cluster Editing with Vertex Splitting} in an attempt to allow for such overlapping clusters\footnote{Preliminary versions of parts of this paper have been presented at ISCO 2018 and IPEC 2023 (see \cite{abu2018cluster} and \cite{AMGSW2023}).}. 
%
%
We show the new problem to be \NP-complete and investigate its parameterized complexity.
We obtain a polynomial kernel using the notion of a critical cliques as introduced by \citet{lin2000phylogenetic} and applied to \textsc{Cluster Editing} by \citet{guo2009more}.

This paper is structured as follows:
In \Cref{sec:prelim}, we give some basic definitions and notation used throughout the paper.
In \Cref{sec:nphardness}, we prove that our problem is \NP-complete, even on graphs of bounded maximum degree.
In \Cref{sec:editseq}, we study the order of operations in an optimal solution 
and \Cref{sec:critcliq} is devoted to critical cliques.
In \Cref{sec:kernel}, we show how to obtain a~$6k$-vertex kernel in linear time. 
\Cref{sec:algo} presents a fixed-parameter tractable algorithm and we conclude
in \Cref{sec:conclusion} with some open problems and future directions.

\section{Preliminaries}
\label{sec:prelim}

For a positive integer~$n$, we use~$[n]$ to denote the set~$\{1,2,\ldots,n\}$ of all positive integers up to~$n$.
All logarithms in this paper use 2 as their base.
We use standard graph-theoretic notation and refer the reader to the textbook by Diestel~\cite{diestel2005graphtheory} for commonly used definitions.
All graphs in this work are simple, unweighted, and undirected.
We denote the open and closed neighborhoods of a vertex~$v$ by~$N(v)$ and~$N[v]$, respectively.
For a subset $V'$ of vertices in a graph~$G$, we denote by $G[V']$ the subgraph of $G$ induced by $V'$.
For an introduction to parameterized complexity, fixed-parameter tractability, and kernelization, we refer the reader to the textbooks by Flum and Grohe~\cite{flum2006parameterizedcomplexity}, Niedermeier~\cite{Niedermeier06}, and Cygan et al.~\cite{CFKLMPPS15}.

The \emph{exponential-time hypothesis} (ETH), formulated by Impagliazzo, Paturi, and Zane~\cite{impagliazzo2001whichproblems}, states that there exists some positive real number $s$ such that \textsc{3-Sat} on $N$ variables and $M$ clauses cannot be solved in $2^{s(N+M)}$ time.





A \emph{cluster graph} is a graph in which the vertex set of each connected component induces a clique.
Equivalently, a graph is a cluster graph if and only if the graph does not have $P_3$ as an induced subgraph.

\subsection*{Problem Definition}
Given a graph~$G=(V,E)$, an edit sequence of length~$k$ is a sequence~$\sigma = (e_1,e_2,\dots,e_k)$ of~$k$ operations, where each~$e_i$ is one of the following operations:
\begin{enumerate}
  \item do nothing,
  \item add an edge to $E$,
  \item delete an edge from $E$, and
  \item \label{icevd:ivs} split a vertex, that is, replace a vertex~$v$ by two vertices~$v_1,v_2$ such that~$N(v_1) \cup N(v_2) = N(v)$.
\end{enumerate}
An example of the splitting operation is given in \cref{fig:incsplit} and we call the two new vertices~$v_1$ and~$v_2$ \emph{copies} of the original vertex~$v$ (and if~$v_1$ or~$v_2$ are further split in the future, then the resulting vertices are also called copies of~$v$).
\begin{figure}[t]
	\centering
    \begin{tikzpicture}
        \node[circle, draw, label=$a$] at (-1,0) (a) {};
        \node[circle, draw, label=$b$] at (0,1) (b) {};
        \node[circle, draw, label=$c$] at (1,0) (c) {};
        \node[circle, draw, label=right:$d$] at (0,-1) (d) {};
        \node[circle, draw, label=above right:$v$] at (0,0) (v) {} edge(a) edge(b) edge(c) edge(d);

        \node at(2.5,0) {\LARGE $\Rightarrow$};
        
        \node[circle, draw, label=$a$] at (4,0) (a2) {};
        \node[circle, draw, label=$b$] at (6,1) (b2) {};
        \node[circle, draw, label=$c$] at (7,0) (c2) {};
        \node[circle, draw, label=right:$d$] at (5,-1) (d2) {};
        \node[circle, draw, label=$v_1$] at (5,0) (v1) {} edge(a2) edge(b2) edge(d2);
        \node[circle, draw, label=below:$v_2$] at (6,0) (v2) {} edge(b2) edge(c2);
    \end{tikzpicture}
    \caption{An illustration of a vertex-split operation. The vertex~$v$ is replaced by~$v_1$ and~$v_2$, with the vertices~$a,c,$ and~$d$ being adjacent to exactly one of the two vertices and~$b$ being adjacent to both.}
    \label{fig:incsplit}
\end{figure}
We denote the graph resulting from applying an edit sequence~$\sigma$ to a graph~$G$ by~$\ver{G}{\sigma}$.
\textsc{Cluster Editing with Vertex Splitting} is then defined as follows.
Given a graph $G$ and an integer $k$, is there an edit sequence $\sigma$ of length~$k$ such that~$\ver{G}{\sigma}$ is a cluster graph?

\section{\NP-hardness}
\label{sec:nphardness}

\noindent
In this section, we show that \textsc{Cluster Editing with Vertex Splitting} is \NP-complete.

\begin{theorem}
	\CEVS{} is \NP-complete. Moreover, assuming the exponential-time hypothesis, there is no~$2^{o(n+m)}$-time or~$2^{o(k)} \cdot \poly(n)$-time algorithm for it.
\end{theorem}

\begin{proof}
Since containment in \NP{} is obvious (non-deterministically guess the sequence of operations and check that the resulting graph is indeed a cluster graph), we focus on the \NP-hardness and present a reduction from \textsc{3-Sat}.
Therein, we will use two gadgets, a \emph{variable gadget} and a \emph{clause gadget}.
The variable gadget is a wheel graph with two (connected) center vertices.
An example of this graph is depicted on the left side of \cref{fig:wheel}.
We call this graph with~$t$ vertices on the outside~$W_t$ and we will only consider instances with~$t \bmod 6 = 0$, that is,~$t=6a$ for some positive integer~$a$.
The clause gadget is a ``crown graph'' as depicted in \cref{fig:crown}(a).

\input{figwheel}

\input{figcrown}

More precisely, for each variable~$x_i$, we construct a variable gadget~$G_i$ which is a~$W_{6a}$ where~$a$ is the number of clauses that contain either~$x_i$ or~$\neg x_i$.
For each clause~$C_j$, we construct a clause gadget~$H_j$ as depicted in \cref{fig:crown}(a), that is, a~$K_5$ with the edges of a triangle removed.
We arbitrarily assign each of the three vertices of degree two in~$H_j$ to one literal in~$C_j$.
Finally, we connect the variable and clause gadgets as follows.
If a variable~$x_i$ appears in a clause~$C_j$, then let~$u$ be the vertex in~$H_j$ assigned to~$x_i$ (or~$\neg x_i$).
Moreover, let~$b$ be the number such that~$C_j$ is the~$b$\textsuperscript{th} clause containing either~$x_i$ or~$\neg x_i$ and let~$c = 6 (b-1)$.
Let the vertices on the outer cycle of~$G_i$ be~$v_1,v_2,\ldots,v_{6a}$.
If~$C_j$ contains the literal~$x_i$, then we add the three edges~$\{u,v_{c+1}\},\{u,v_{c+2}\},\{u,v_{c+3}\}$.
If~$C_j$ contains the literal~$\neg x_i$, then we add the three edges~$\{u,v_{c+2}\},\{u,v_{c+3}\},\{u,v_{c+4}\}$.
To complete the reduction, we set~$k = 35M - 2N$, where~$M$ is the number of clauses and~$N$ is the number of variables.

We next show that the reduction is correct, that is, the constructed instance of \CEVS{} is a yes-instance if and only if the original formula~$\phi$ of 3-\textsc{Sat} is satisfiable.
To this end, first assume that~$\phi$ is satisfiable and let~$\beta$ be a satisfying assignment.
For each variable~$x_i$, we will partition~$G_i$ into~$K_5$'s as follows.
Let~$a$ be the value such that~$G_i$ is isomorphic to~$W_{6a}$.
If~$\beta$ sets~$x_i$ to true, then we remove the edge~$\{v_{3j},v_{3j+1}\}$ and add the edge~$\{v_{3j+1},v_{3j+3}\}$ for each integer~$1 \leq j \leq 2a$ (where values larger than~$6a$ are taken modulo~$6a$).
If~$\beta$ sets~$x_i$ to false, then we remove the edge~$\{v_{3j+1},v_{3j+2}\}$ and add the edge~$\{v_{3j+2},v_{3j+4}\}$ for each~$1 \leq j \leq 2a$.
Moreover, we split the two center vertices~$2a-1$ times.
In total, we use~$8a-2$ modifications to transform~$G_i$ into a collection of~$K_5$'s.
Since each clause contains exactly three literals and we add six vertices for each variable appearance, the sum of lengths of cycles in all variable gadgets combined is~$18M$.
Hence, in all variable gadgets combined, we perform~$24M - 2N$ modifications.

Next, we modify the crown graphs.
To this end, let~$C_j$ be a clause and let~$H_j$ be the constructed clause gadget.
Since~$\beta$ is a satisfying assignment, at least one variable appearing in~$C_j$ satisfies it.
If multiple such variables exist, then we pick one arbitrarily.
Let~$x_i$ be the selected variable and let~$u$ be the vertex in~$H_j$ assigned to~$x_i$.
We first turn~$H_j$ into a~$K_4$ and an isolated vertex by removing the two edges incident to~$u$ in~$H_j$ and add the missing edge between the two vertices assigned to different variables.
Finally, we look at the edges between variable gadgets and clause gadgets.
For the vertex~$u$, note that by construction the three vertices that~$u$ is adjacent to in~$G_i$ already belong to a~$K_5$ and hence we can add two edges to (copies of) the two centers of the variable gadget to form a~$K_6$.
For the two other vertices in~$H_j$ that have edges to vertices in variable gadgets, we remove all three such edges, that is, six edges per clause.
Hence, we use~$3+2+6 = 11$ modifications for each clause.
Since the total number of modifications is~$35M - 2N$ and the resulting graph is a collection of~$K_4$'s, $K_5$'s, and~$K_6$'s, the constructed instance of \CEVS{} is a yes-instance.

For the other direction, suppose the constructed instance of \CEVS{} is a yes-instance.
We first show that~$24M - 2N$ modifications are necessary to transform all variable gadgets into cluster graphs and that this bound can only be achieved if each time exactly three consecutive vertices on the cycles are contained in the same~$K_5$.
To this end, consider any variable gadget~$G_i$.
By construction,~$G_i$ is isomorphic to~$W_{6a}$ for some integer~$a$.
By the counting argument from above, we show that at least~$8a-2$ modifications are necessary.
Note first that some edge in the cycle has to be removed or some vertex on the cycle has to be split as otherwise any solution would contain a clique with all vertices in the cycle and this would require at least~$18a^2 - 9a > 8a-2$ edge additions (since the degree of each of the~$6a$ vertices in the cycle would need to increase from~$2$ to~$6a-1$).
We next analyze how many modifications are necessary to separate~$b$ vertices from the outer cycle into a clique.
We require at least two modifications for the center vertices (either splitting them or removing the edges between them and the first vertex that we want to separate) and one operation to separate the cycle on the other end (either splitting a vertex or removing an edge of the cycle).
For~$b \in \{1,2\}$ these operations are enough.
For~$b \geq 3$, we need to add~$\binom{b}{2}-(b-1)$ edges (all edges in a clique of size~$b$ minus the already existing edges of a path on~$b$ vertices).
Note that the ``average cost'' per separated vertex (number of operations divided by~$b$) is minimized (only) with~$b = 3$ with a cost of~$4$ for three vertices.
Hence, to separate all but~$c$ vertices from the cycle, we require at least~$4(6a-c)/3$ operations.
The cost for making the remaining~$c$~vertices into a clique requires again $\binom{c}{2}-(c-1)$~edge additions.
Analogously, the optimal solution is to have~$c=3$ with just a single edge addition.
Thus, the minimum number of required operations is at least~$4(6a-3)/3 + 2 = 8a - 2$ (where the $+2$ comes from the initial edge removal and the final edge addition between the last~$c=3$ vertices) and this value can only be reached by partitioning the cycle into triples which each form a~$K_5$ with the two center vertices.
Note that it is always preferable to delete an edge on the outer cycle and not split one of the two incident edges as splitting a vertex increases the number of vertices on the cycle and thus invokes a higher overall cost.
Next, we analyze the clause gadget and the edges between the different gadgets.
We start with the latter.
Let~$u$ be a vertex in a clause gadget~$H_j$ with (three) incident edges to some variable gadget.
The only way to not use at least three operations to deal with the three edges is if~$u$ is an isolated vertex or if the three neighbors do not have two more neighbors in the current solution.
In the former case, we can (possibly) add the two edges between~$u$ and the two centers of the respective variable gadgets to build a~$K_6$.
In the latter case, we have used at least three operations more in the variable gadget than intended (either by removing edges between neighbors of~$u$ and the center vertices or by splitting all neighbors of~$u$).
Since each vertex in a variable gadget is only adjacent to at most one vertex in a clause gadget, this cannot lead to an overall reduction in the number of operations and we can therefore ignore this latter case.

We are now in a position to argue that at least eleven modifications are necessary for each clause gadget.
First, note that at least three operations are required to transform a crown into a cluster graph.
Possible ways of achieving this are depicted in \cref{fig:crown}.
In each of these possibilities, at most one vertex becomes an isolated vertex.
To make two vertices independent, at least four operations are required and for three isolated vertices, at least five operations are required.
As shown above, at least two operations are required for each isolated vertex with edges to variable gadgets and at least three operations are required for non-isolated vertices with edges to variable gadgets.
Thus, at least eleven operations are required for each clause gadget and eleven operations are sufficient if and only if the three vertices incident to one of the vertices in~$H_j$ belong to the same~$K_5$ in the variable gadget.

By the argument above, at least~$24M - 2N + 11M = k$ operations are necessary and since the constructed instance is a yes-instance, there is a way to cover all variable gadgets with~$K_5$'s such that for each clause there is at least one vertex whose three neighbors in a variable gadget belong to the same~$K_5$.
Let~$C_j$ be a clause, let~$u$ be a vertex with all three neighbors in the same~$K_5$, and let~$x_i$ be the variable corresponding to this variable gadget.
If~$x_i$ appears positively in~$C_j$, then~$v_{3i+1},v_{3i+2},$ and~$v_{3i+3}$ belong to the same~$K_5$ for each~$i$ and we set~$x_i$ to true.
If~$x_i$ appears negatively in~$C_j$, then~$v_{3i+2},v_{3i+3},$ and~$v_{3i+4}$ belong to the same~$K_5$ for each~$i$ and we set~$x_i$ to false.
Note that we never set a variable to both true and false in this way.
We set all remaining variables arbitrarily to true or false.
By construction, the variable~$x_i$ satisfies~$C_j$ and since we do the same for all clauses, all clauses are satisfied, that is, the original formula~$\phi$ is satisfiable.
Thus, the constructed instance is equivalent to the original instance of \textsc{3-Sat}.

Since the reduction can clearly be computed in polynomial time, this concludes the proof for the \NP-hardness.
For the ETH-based hardness, observe that~$k,n,m \in O(N + M)$.
This implies that there are no~$2^{o(n+m)}$-time or~$2^{o(k)} \cdot \poly(n)$-time algorithms for \CEVS{} unless the ETH fails \cite{impagliazzo2001whichproblems}.
\ifarxiv\else\qed\fi
\end{proof}

In contrast to the reduction for \textsc{Cluster Editing}~\cite{komusiewicz2012clusterediting}, our reduction does not produce instances with constant maximum degree.
We instead observe that in our reduction, the maximum degree of the produced instances depends only on the maximum number of times a variable appears in a clause.
Combining this with the fact that \textsc{3-Sat} remains \NP-hard when restricted to instances where each variable appears in at most four clauses~\cite{tovey1984asimplified}, we obtain the following corollary:

\begin{corollary}
    \CEVS{} remains \NP-hard on graphs with maximum degree 24.
\end{corollary}

\section{The Edit-Sequence Approach}
\label{sec:editseq}

In this section, we show that we can always assume that a solution to \CEVS{} has a specific structure, that is, it first adds edges, then removes edges, and finally splits vertices.
We mention that removing edges can also be moved to the front or to the back and that we do not use the fact that we want to reach a cluster graph at any point.
The statement therefore holds for any graph-modification problem which only adds edges, deletes edges, and splits vertices.

To start, we say that two edit sequences $\sigma$ and $\sigma'$ are \emph{equivalent} if $\ver{G}{\sigma}$ and $\ver{G}{\sigma'}$ are isomorphic.
We show first that all vertex splittings can be moved to the back of the edit sequence.
\begin{lemma}
\label{3}
For any edit sequence $\sigma = (e_1,e_2,\dots,e_i,e_{i+1},\dots,e_k)$ where~$e_i$ is a vertex splitting and~$e_{i+1}$ is an edge addition, an edge deletion, or a do-nothing operation, there is an equivalent edit sequence~$\sigma' = (e_1,e_2,\dots,e_{i-1},e'_i,e'_{i+1},e_{i+2},\dots,e_k)$ of the same length where~$e'_i$ is an edge addition, an edge deletion, or a do-nothing operation and~$e_{i+1}'$ is a vertex splitting.
\end{lemma}

\begin{proof}
If~$e_{i+1}$ is a do-nothing operation or the edge added or removed by it is not incident to one of the vertices introduced by the vertex split~$e_i$, then the edit sequence~$\sigma'$ where~$e'_{i} = e_{i+1}$ and~$e'_{i+1} = e_i$ is equivalent to~$\sigma$ and of the same length.
So assume without loss of generality that~$e_i$ splits vertex~$v$ into~$v_1$ and~$v_2$ and~$e_{i+1}$ adds or deletes the edge~$\{v_1,w\}$ for some vertex~$w$.
If~$e_{i+1}$ adds the edge~$\{v_1,w\}$ and the edge~$\{v,w\}$ exists after performing the edit subsequence~${\sigma_1 = (e_1,e_2,\ldots,e_{i-1})}$, then we set~$e'_{i}$ to be the do-nothing operation.
If~$e_{i+1}$ adds the edge~$\{v_1,w\}$ and the edge~$\{v,w\}$ does not exist after performing~$\sigma_1$, then we let~$e'_{i}$ add the edge~$\{v,w\}$.
In both cases, we let~$e'_{i+1} = e_i$ with the modification that~$v_1$ is also adjacent to~$w$.
Note that this is possible as~$v$ is adjacent to~$w$ after performing the edit sequence~$(e_1,e_2,\ldots,e_{i-1},e'_i)$.
If~$e_{i+1}$ deletes the edge~$\{v_1,w\}$, then we know that the edge~$\{v,w\}$ exists after performing~$\sigma_1$ and we can assume without loss of generality that the edge~$\{v_2,w\}$ does not exists after performing the edit sequence~$(e_1,e_2,\ldots,e_{i-1},e'_i,e'_{i+1})$.
Hence, we let~$e'_i$ remove the edge~$\{v,w\}$ and let~$e'_{i+1} = e_i$ with the modification that~$v_1$ is no longer adjacent to~$w$.
In all cases, the graphs reached after performing the edit sequences~$(e_1,e_2,\ldots,e_i,e_{i+1})$ and~$(e_1,e_2,\ldots,e_{i-1},e'_i,e'_{i+1})$ are identical and hence performing the edit sequence~$\sigma^* = (e_{i+2},e_{i+3},\ldots,e_k)$ afterwards shows that~$\sigma$ and~$\sigma'$ are equivalent (and they are of the same length).
\ifarxiv\else\qed\fi
\end{proof}

We show next that moving edge additions to the front results in an equivalent edit sequence. 

\begin{lemma}
\label{1}
For any edit sequence $\sigma = (e_1,e_2,\dots,e_i,e_{i+1},\dots,e_k)$ where $e_i$ is an edge deletion and $e_{i+1}$ is an edge addition, at least one of the edit sequences $\sigma' = (e_1,e_2,\dots,e_{i-1},e'_i,e'_{i+1},e_{i+2},\dots,e_k)$ where $e'_i = e_{i+1}$ and~$e_{i+1}' = e_i$ or both are do-nothing operations is of the same length and equivalent to $\sigma$.
\end{lemma}

\begin{proof}
Note that if~$e_{i+1}$ adds the edge that~$e_{i}$ removed, then replacing these two (consecutive) operations with do-nothing operations results in the same graph.
If~$e_{i+1}$ adds a different edge than~$e_{i}$ removed, then the graphs reached after the edit subsequences~$\sigma_1 = (e_1,e_2,\ldots,e_{i},e_{i+1})$ and~${\sigma_2 = (e_1,e_2,\ldots,e_{i+1},e_{i})}$ are identical.

Hence, performing the edit sequence~$\sigma^* = (e_{i+2},e_{i+3},\ldots,e_k)$ afterwards results in isomorphic graphs for both starting edit sequences.
Note that performing~$\sigma_1$ first and then~$\sigma^*$ is equivalent to performing~$\sigma$ and performing~$\sigma_2$ first and~$\sigma^*$ afterwards is equivalent to performing~$\sigma'$ where~$e'_i = e_{i+1}$ and~$e_{i+1}' = e_i$.
This concludes the proof.
\ifarxiv\else\qed\fi
\end{proof}

We can easily deduce the following theorem from the two above lemmata.

\begin{theorem}
\label{thm:add}
For any edit sequence $\sigma = (e_1,e_2,\ldots,e_k)$, there is an edit sequence~${\sigma'  = (e'_1,e'_2,\ldots,e'_{k'})}$ such that
\begin{enumerate}
\item $k' \leq k$,
\item $\sigma$ and~$\sigma'$ are equivalent,
\item \label{canmin:dn} $e'_i$ is not a do-nothing operations for any~$i \in [k']$.
\item \label{canmin:eaed} if $e'_i$ is an edge addition and $e'_j$ is an edge deletion or a vertex splitting, then $i < j$,
\item \label{canmin:edvd} if $e'_i$ is an edge deletion and $e'_j$ is a vertex splitting, then $i < j$, and
\end{enumerate}
\end{theorem}

\begin{proof}
    Let~$\sigma = (e_1,e_2,\ldots,e_k)$ be any edit sequence.
    If~$e_i$ is a do-nothing operation for some~$i\in [k]$, then the edit sequence~$(e_1,e_2,\ldots,e_{i-1},e_{i+1},\ldots,e_k)$ is equivalent and shorter.
    Hence, we can assume that~$\sigma$ does not contain any do-nothing operations.
    If there exists a vertex-splitting operation~$e_i$ and an operation~$e_j$ with~$j>0$ such that~$e_j$ is not a vertex-splitting operation, then let~$i'$ be the last index of a vertex-splitting operation such that~$e_{i'+1}$ is not a vertex-splitting operation.
    By \cref{3}, we can modify~$\sigma$ into an equivalent edit sequence~$(e'_1,e'_2,\ldots,e'_k)$ where~$e'_j = e_j$ for all~$j \in [k] \setminus \{i',i'+1\}$ and~$e_{i'+1}$ is a vertex-splitting operation and~$e_{i'}$ is not.
    If~$e'_{i'}$ is a do-noting operation, we can again remove it.
    Performing this procedure repeatedly until no longer applicable results in an edit sequence~$\sigma_1$ that is equivalent to~$\sigma$, does not contain any do-nothing operations and all edge-addition and edge-deletion operations come before all vertex-splitting operations.
    If~$\sigma_1$~performs an edge-deletion operation~$e_i$ before an edge-addition operation~$e_j$ ($i<j$), then there is also an index~$i'$ such that~$e_{i'}$ is an edge-deletion operation and~$e_{i'+1}$ is an edge-addition operation.
    We can then use \cref{1} to obtain a new sequence in which the number of index pairs~$(i,j)$ with~$i < j$,~$e_i$ is an edge-deletion operation, and~$e_{i}$ is an edge-addition operation is reduced by at least one.
    If the application of \cref{1} introduces a do-nothing operation, then we again remove it.
    Repeatedly applying \cref{1} finally results in an equivalent sequence~$\sigma'$ where the number of mentioned index pairs is zero, that is, all edge-addition operation come before all edge-deletion operations.
    Note that~$\sigma'$ now satisfies all requirements of the theorem statement.
\end{proof}

We refer to an edit sequence satisfying the statement of \cref{thm:add} as an \emph{edit sequence in standard form}.

\section{Critical Cliques}
\label{sec:critcliq}

Originally introduced by \citet{lin2000phylogenetic}, critical cliques provide a useful tool in understanding the clique structure in graphs. 
\begin{definition}
  A \emph{critical clique} is a subset of vertices $C$
  that is maximal with the properties that
  \begin{enumerate}
  \item $C$ is a clique
  \item there exists $U \subseteq V(G)$ s.t.~$N[v] = U$ for all
    $v \in C$.
  \end{enumerate}
\end{definition}

\noindent
Equivalently, two vertices~$u$ and~$v$ belong to the same critical clique if and only if they are true twins, that is,~$N[u] = N[v]$.
Hence, each vertex~$v$ appears in exactly one critical clique.
The \emph{critical clique quotient graph}~$\mathcal C$ of~$G$ contains a node for each critical clique in~$G$ and two nodes are adjacent if and only if the two respective critical cliques $C_1$ and $C_2$ are adjacent, that is, there is an edge between each vertex in~$C_1$ and each vertex in~$C_2$.
Note that by the definition of critical cliques, this is equivalent to the condition that at least one edge~$\{u,v\}$ with~$u \in C_1$ and~$v \in C_2$ exists.
To avoid confusion, we will call the vertices in~$\mathcal{C}$ nodes and in~$G$ vertices.

The following lemma is adapted from Lemma 1 by \citet{guo2009more} with a careful restatement in the context of our new problem.\footnote{We mention in passing that we claimed a slightly stronger lemma in a previous version of this paper. \citet{FDH+23} observed that the stronger version does not hold and conjectured that this weaker version is true. We confirm this here.}

\begin{lemma}[Critical clique lemma]
\label{lem:ccl}
  Let $(G, k)$ be a yes-instance of \CEVS.
  Then, there exists a solution~$\sigma$ of length at most~$k$ such that for each critical clique~$C_i$ in~$G$ and each clique~$S_j$ in $\ver{G}{\sigma}$, either~$S_j$ contains exactly one copy of each vertex in~$C_i$ or~$S_j$ does not contain any copy of a vertex in~$C_i$.
\end{lemma}
\begin{proof}
  Let $\hat\sigma$ be an optimal solution for~$(G,k)$ in standard form.
  %
  For each critical clique $C_i$, we select a representative vertex
  $r_i \in C_i$ by picking any vertex in $C_i$ with fewest appearances
  in $\hat\sigma$.
  If it holds for each critical clique~$C_i$ in~$G$ and each clique~$S_j$ in the resulting cluster graph~$\ver{G}{\hat\sigma}$ that~$S_j$ contains exactly one copy of each vertex in~$C_i$ or~$S_j$ does not contain any copy of a vertex in~$C_i$, then~$\hat\sigma$ satisfies all requirements of the lemma statement and we are done.
  Otherwise, there exists a clique $S_j$ in~$\ver{G}{\hat\sigma}$ which contains two copies of some vertex~$v$ or there exists a critical clique $C_i$ and two vertices~$u,v \in C_i$ such that~$S_j$ contains a copy of~$u$ but no copy of~$v$.
  In the former case, note that removing any operations involving one of the two copies results in a solution of strictly shorter length, contradicting the fact that~$\hat\sigma$ was an optimal solution.
  In the latter case, there also exists such a pair of vertices where one of the two vertices is~$r_i$.
  Let~$w$ be the other vertex.
  
  We find a new optimal solution by removing all operations involving~$w$ and copying all operations including $r_i$ and replacing $r_i$ by $w$ in the copy.
  For the sake of notational convenience, we say that that if some operation involves the~$j$\textsuperscript{th} copy of~$r_i$, then the very next operation will be the copy and will use the $j$\textsuperscript{th} copy of~$w$.\footnote{We only point out here that since we removed any operations involving~$w$, the original edge between~$r_i$ and~$w$ is not removed. Moreover when splitting another vertex (not $r_i$ or $w$), then we treat each copy of $w$ the same as the corresponding copy of~$r_i$. When splitting the $j$\textsuperscript{th} copy of $r_i$ to create the~$j'$\textsuperscript{th} and~$j''$\textsuperscript{th} copies, then we keep both new vertices adjacent to the $j$\textsuperscript{th} copy of $w$ and when splitting the $j$\textsuperscript{th} copy of $w$, then we make the $j'$\textsuperscript{th} copy of $w$ adjacent to the $j'$\textsuperscript{th} copy of $r_i$ and the $j''$\textsuperscript{th} copy of $w$ adjacent to the $j''$\textsuperscript{th} copy of $r_i$. Note that each copy of $w$ is adjacent to the respective copy of $r_i$ and not adjacent to any other copy of $r_i$ (and vice versa).}

  Since $w$ is involved in at least as many operations as $r_i$ (recall that $r_i$ was picked as having fewest appearances in $\hat\sigma$), the resulting sequence $\sigma'$ will be at most as long as~$\hat\sigma$.
  We next show that~$\sigma'$ is also a solution.
  If the resulting graph~$\ver{G}{\sigma'}$ is not a cluster graph, then it contains an induced~$P_3$.
  Let~$X$ be an induced~$P_3$ in~$\ver{G}{\sigma'}$.
  Since we only modified operations for $w$, some copy of~$w$ has to be part of $X$.
  Let~$j$ be the index such that~$X$ contains the $j$\textsuperscript{th} copy of~$w$.
  If~$X$ contains another copy of~$w$, then the two copies are the two ends.
  The center cannot be a copy of~$r_i$ as each copy of~$w$ is only adjacent to the corresponding copy of~$r_i$ and vice versa.
  Hence, we have another induced $P_3$
  where we replace the two copies of~$w$ by their respective copies of~$r_i$.
  This contradicts that $\hat\sigma$ is a solution.

  So assume that~$X$ contains only the $j$\textsuperscript{th} copy of~$w$.
  If~$X$ does not contain the $j$\textsuperscript{th} copy of~$r_i$, then the graph
  contains also a~$P_3$ where we replace the~$j$\textsuperscript{th} copy~$w$ by the~$j$\textsuperscript{th} copy of~$r_i$.
  This again contradicts the assumption that $\hat\sigma$ is a solution.
  Thus, we may assume that $X$ contains both the $j$\textsuperscript{th} copy of $r_i$ and the~$j$\textsuperscript{th} copy of~$w$.
  Since these two vertices are adjacent, one of the two vertices is the
  center of $X$.
  However, all other vertices have either both or none of the two
  vertices as neighbors.
  This contradicts that $X$ exists.
  Hence, $\sigma'$ is also a solution.

  Note that the number of vertices that behave differently than their representative is one smaller in~$\sigma'$ compared to~$\hat\sigma$.
  Hence, repeating the above procedure at most~$n$ times results results in an optimal solution~$\sigma$ as stated in the lemma.
  \ifarxiv\else\qed\fi
\end{proof}

In the following two sections, we will use \cref{lem:ccl} to develop a linear-vertex kernel and a~$2^{O(k \log k)}(n+m)$-time algorithm.

\section{A
\texorpdfstring{$6k$}{6k}-
vertex kernel
}
\label{sec:kernel}

In this section, we prove a problem kernel for \CEVS{} with at most~$6k$ vertices based on the critical clique lemma (\Cref{lem:ccl}) and a similar kernel for \textsc{Cluster Editing} by \citet{guo2009more}.
To this end, we propose three reduction rules, prove that they are safe\footnote{A reduction rule is safe if its application results in an equivalent instance.}, that they can be performed exhaustively in linear time, and that their application gives a kernel as required.
We start with the simplest one.

\begin{rul}
\label{kern:rul1}
Remove all isolated cliques.
\end{rul}
\begin{lemma}
Reduction Rule \ref{kern:rul1} is safe. 
\end{lemma}

\begin{proof}
Notice that for any optimal solution~$\sigma$, each clique in~$\ver{G}{\sigma}$ only contains copies of vertices of one connected component in~$G$.
Hence, if a clique contains a copy of a vertex in an isolated clique~$K$ in~$G$, then it only contains copies of vertices in~$K$.
Since~$K$ by definition does not require any operations to be transformed into a clique with no outgoing edges, removing~$K$ results in an equivalent instance of \CEVS.
\ifarxiv\else\qed\fi
\end{proof}

We next bound the size of each critical clique in terms of the sizes of adjacent critical cliques.

\begin{rul}
\label{kern:rul4}
If there is a critical clique~$K$ such that~$|K| > |\bigcup_{C \in N(K)} C| + 1$, then reduce the size of~$K$ to~$|\bigcup_{C \in N(K)} C| + 1$.
\end{rul}
\begin{lemma}
\label{rul:2}
Reduction Rule \ref{kern:rul4} is safe.
\end{lemma}
\begin{proof}
Let~$K$ be a critical clique with~$|K| > |\bigcup_{C \in N(K)} C|$ and let~$\sigma$ be an optimal solution.
We show that~$\sigma$ does not split any vertex corresponding to~$K$ and does not add or delete any edge incident to such a vertex. 
By \cref{lem:ccl}, we may assume that all cliques in~$\ver{G}{\sigma}$ contain no or exactly one copy of vertices in~$K$.
Let~$H$ be a clique in~$\ver{G}{\sigma}$ that contains exactly one copy of each vertex corresponding to~$K$.
Let~$A$ be the set of nodes in~$N(K)$ in the critical clique quotient graph~$\mathcal{C}$ of~$G$ whose corresponding vertices have a copy in~$H$.
Analogously, let~$B$ be the set of nodes not in~$N[K]$ in~$\mathcal{C}$ whose corresponding vertices have a copy in~$H$.
Note first that we can assume~$B = \emptyset$ as otherwise~$\sigma$ adds at least~$|K|$ edges between a vertex corresponding to a node in~$B$ and all vertices in~$K$.
Splitting all vertices corresponding to nodes in~$A$ instead (and therefore splitting the clique~$H$ into two cliques, one containing a copy of each vertex corresponding to a node in~$K \cup A$ and the other containing a copy of each vertex corresponding to a node in~$A \cup B$) results in another cluster graph reached by a shorter edit sequence as~$|A| \leq |\bigcup_{C \in N(K)} C| < |K|$.
This contradicts the fact that~$\sigma$ is an optimal solution.
We next show that~$A = N(K)$.
Assume towards a contradiction that~$A \neq N(K)$.
Then, there exists a node~$v \in N(K) \setminus A$.
Modifying~$\sigma$ to split each vertex corresponding to~$v$ once, add all missing edges between vertices corresponding to~$v$ and vertices corresponding to a node in~$A$, and removing all edge removals between vertices corresponding to~$K$ and~$v$ results in another cluster graph.
Moreover, the newly acquired edit sequence is in fact shorter as~$|v| + |v| \cdot |\bigcup_{C\in A}C| \leq |v|\cdot (|\bigcup_{C\in N(K) \setminus \{v\}}C| + 1) \leq |v| \cdot (|\bigcup_{C\in N(K)}C|) < |v|\cdot |K|$.
This again contradicts the fact that~$\sigma$ is optimal.
We have shown that~$A = N(K)$ and~$B = \emptyset$.
Since~$H$ contains a copy of each vertex adjacent to a vertex in~$K$, we can assume without loss of generality that no vertex in~$K$ is split and no edge incident to such a vertex is added or deleted.
Since the above argument holds as long as~$|K| > |\bigcup_{C \in N(K)} C|$, we can reduce the size of~$K$ to~$|\bigcup_{C \in N(K)} C|+1$ and still have an equivalent instance of \CEVS.
\ifarxiv\else\qed\fi
\end{proof}

We next show that if more than~$6k$ vertices are left after performing Reduction Rules 1 and 2 exhaustively, then we have a no-instance. 
\begin{lemma}
\label{lem:cclnum}
If there is a solution to \CEVS{} on $(G,k)$, then there are at most~$6k$ vertices and~$4k$ critical cliques left after performing Reduction Rules 1 and 2 exhaustively.
\end{lemma}

\begin{proof}
We follow an approach similar to that taken by~\citet{guo2009more}. 
Let~$\sigma$ be an optimal solution of \CEVS{} that satisfies \cref{lem:ccl}.
Let~$G'$ be the graph obtained from~$G$ after applying Reduction Rules 1 and 2 exhaustively.
Partition the node set of the critical clique quotient graph~$\mathcal{C}$ of $G'$ into 4 sets~$W,X,Y,$ and~$Z$ as follows.
Let~$W$ be the set of nodes whose corresponding vertices are the endpoint of some edge added by $\sigma$.
Let~$X$ be the subset of nodes not contained in~$W$ whose corresponding vertices are the endpoint of some edge deleted by $\sigma$.
Let~$Y$ be the subset of nodes not in~$W$ or~$X$ whose corresponding vertices are split by $\sigma$.
Finally, let~$Z$ be all other nodes in $\mathcal{C}$.
As each vertex corresponding to a node in $W$, $X$, and $Y$ is affected by some operation in $\sigma$ and each operation can affect at most $2$ vertices, it holds that~$|\bigcup_{C \in W\cup X\cup Y} C| \leq 2|\sigma| \leq 2k$.

Let us now consider~$Z$.
Assume towards a contradiction that there is a clique~$H$ in~$\ver{G}{\sigma}$ that contains the vertices corresponding to a node in~$Z$ but no vertex corresponding to a node in~$W \cup X \cup Y$ or~$C$ contains vertices corresponding to two nodes in~$Z$.
In the former case, let~$v \in Z$ be a node whose corresponding vertices are contained in~$H$.
By definition of~$Z$, the vertices corresponding to~$v$ are adjacent to all vertices in~$H$ (or the vertices whose copies are in~$H$).
Hence, none of these vertices correspond to a node in~$W \cup X \cup Y$.
Since Reduction Rule~1 removed all isolated critical cliques,~$v$ is not an isolated node and there is therefore a second node~$u \in Z$ whose corresponding vertices are contained in~$H$.
Hence, we are in the second case where~$H$ contains the vertices of at least two critical cliques~$u,v \in Z$.
By definition of~$Z$, the vertices corresponding to~$u$ and~$v$ are adjacent to all vertices in~$H$ (or the vertices whose copies are in~$H$) and not adjacent to any other vertices.
Moreover, since~$H$ is a clique and no edges incident to vertices corresponding to~$u$ or~$v$ were added by~$\sigma$, all vertices corresponding to~$u$ and~$v$ are pairwise adjacent in~$G$.
Note that this means that~$u$ and~$v$ are actually the same critical clique, a contradiction.
Hence, each clique~$H$ in~$\ver{G}{\sigma}$ contains vertices corresponding to at most one node in~$Z$ and if it does, then it contains vertices of at least one node in~$W \cup X \cup Y$.
This also means that the number of nodes in~$Z$ is upper bounded by~$|W \cup X \cup Y| \leq 2k$.
Thus,~$G'$ contains at most~$|W \cup X \cup Y \cup Z| \leq 4k$~critical cliques. 

Consider now the graph~$\ver{G}{\sigma}$.
The number of vertices that are copies of vertices corresponding to nodes in~$W \cup X \cup Y$ is at most~$2k$.
By definition of~$Z$, the vertices corresponding to nodes in~$Z$ are not split during~$\sigma$.
Hence, the number of vertices in~$\ver{G}{\sigma}$ that are copies of vertices corresponding to nodes in~$Z$ is the same as the number of vertices in~$G'$ that correspond to nodes in~$Z$.
Assume towards a contradiction that this number is more than~$4k$.
Then, there exists by the pigeonhole principle a clique~$H$ in~$\ver{G}{\sigma}$ and an integer~$\ell$ such that~$H$ contains~$\ell$ vertices which are copies of vertices corresponding to nodes in~$W \cup X \cup Y$ and at least~$2\ell+1$ vertices that are (copies of vertices) corresponding to nodes in~$Z$.
Let~$A$ be the set of vertices in~$G'$ which correspond to a node in~$W \cup X \cup Y$ and have a copy in~$H$ and let~$B$ be the set of vertices in~$G'$ which correspond to a node in~$Z$ and are contained in~$H$.
By definition of~$Z$, all vertices in~$B$ are adjacent to all vertices in~$A$ in~$G'$.
Moreover, as shown above, all vertices in~$B$ belong to the same critical clique~$C$ in~$G'$.
Hence,~$|C| \geq 2|\bigcup_{C' \in N(C)} C'| + 1 \geq |\bigcup_{C' \in N(C)} C'| + 2$.
This contradicts the fact that~$C'$ was reduced with respect to Reduction Rule 2.
Thus, the number of vertices in~$G'$ that correspond to a node in~$Z$ is at most~$4k$ and the total number of vertices in~$G'$ is at most~$6k$.
\ifarxiv\else\qed\fi
\end{proof}
The above lemma can be converted into the following reduction rule.
Note that a~$C_4$ is a cycle on four vertices and a~$C_4$ cannot be transformed into a cluster graph with a single operation.

\begin{rul}
\label{kern:rul3}
    If there are more than $6k$ vertices or~$4k$ critical cliques left after applying Reduction Rules 1 and 2 exhaustively, then reduce the graph to a $C_4$ and set $k=1$.
\end{rul}

\noindent
Based on the previous three reduction rules, it is easy to derive a problem kernel with~$6k$ vertices in linear time.

\begin{theorem}
\label{thm:kernel}
One can compute a kernel with at most~$6k$ vertices and~$4k$ critical cliques for \CEVS{} in linear time.
\end{theorem}
\begin{proof}
Computing the critical clique quotient graph~$\mathcal{C}$ of~$G$ takes linear time \cite{lin2000phylogenetic}.
Applying Reduction Rule 1 exhaustively also takes linear time and this removes all isolated nodes in~$\mathcal{C}$.
Applying Reduction Rule 2 exhaustively takes linear time as shown next.
First, we can sort all critical cliques by their size in linear time using bucket sort.
Applying Reduction Rule 2 to a critical clique~$C$ takes~$\deg(C)$ time.
By the handshaking lemma, this procedure takes time linear in the number of edges in~$\mathcal{C}$ which is upper-bounded by the number of edges in the input graph.
Note that if we iterate over the critical cliques in increasing order of size, then an application of Reduction Rule 2 can never affect previously considered critical cliques.
This is due to the fact that an application of Reduction Rule 2 does only depend on adjacent critical cliques.
Since, whenever we reduce the size of a critical clique, we only reduce it to a size larger than all of its adjacent critical cliques, this can never lead to a situation where we can reduce a critical clique that was initially smaller than the current critical clique.
Hence, applying Reduction Rule 2 exhaustively takes linear time.
Afterwards, we can compute the critical clique quotient graph~$\mathcal{C}'$ of the resulting graph~$G'$ and count the number of vertices in~$G'$ and~$\mathcal{C}'$ in linear time.
If the number of vertices in~$G$ is at most~$6k$ and the number of vertices in~$\mathcal{C}'$ is at most~$4k$, then we are done.
Otherwise, we apply Reduction Rule 3.
This is correct by \cref{lem:cclnum}, can be performed in constant time, and the resulting graph has~$4 \leq 6k'$ vertices and~$4 \leq 4k'$ critical cliques, where~$k'=1$ is the newly set parameter.
This concludes the proof.
\ifarxiv\else\qed\fi
\end{proof}

We leave it as an open problem whether the size of the kernel (especially the number of edges) can be improved and mention that there is a~$2k$-vertex kernel for \textsc{Cluster Editing}~\cite{ChenMeng}.


\section{An FPT algorithm}
\label{sec:algo}

The result in \cref{thm:kernel} implies that \CEVS{} is fixed-parameter tractable. 
By \cref{lem:ccl}, we can assume that all vertices in the same critical clique belong to the same clique in a solution.
It is easy to see that the final solution consists of at most~$2k$ cliques and guessing these for each of the at most~$4k$~critical cliques takes~$O((2^{2k})^{4k}) = O(2^{8k^2})$ time.
Checking whether a given solution can be reached in~$k$~operations takes~$O(k^2)$ time and combined with the time for computing the kernel, this results in a running time in~$O(2^{8k^2}k^2+n+m)$.
This is, however, far from optimal as shown next.

\begin{theorem}
    \CEVS{} can be solved in~$O(2^{9k \log k} + n + m)$~time.
\end{theorem}

\begin{proof}
    First, we compute the critical clique of each vertex and the critical clique quotient graph~$\mathcal{C}$ of $G$ in linear time \cite{lin2000phylogenetic}.
    Next, we compute the kernel from \cref{thm:kernel} in linear time.
    Note that~$\mathcal{C}$ contains at most~$4k$~vertices.
    By \cref{lem:ccl}, we can also assume that all vertices in a critical clique belong to the same clique in the graph~$\ver{G}{\sigma}$ reached after performing an optimal solution~$\sigma$.
    Let~$\mathcal{X} = \{S_1,S_2,\ldots,S_\ell\}$ be the set of cliques in~$\ver{G}{\sigma}$.
    Note that~$\mathcal{X}$ contains~$\ell \leq 2k$ cliques as each operation can complete at most two cliques of the solution (removing an edge between two cliques or splitting a vertex contained in both cliques) and Reduction Rule 1 removed all isolated cliques (cliques that can be completed without an operation).
    Hence, if there are more than~$2k$ cliques in the solution, then we cannot reach the solution with~$k$ operations.
    To streamline the following argumentation, we will cover the nodes in~$\mathcal{C}$ by cliques~$S_1,S_2,\ldots,S_{\ell=2k}$ and assume that an optimal solution contains exactly~$2k$ cliques by allowing some of the cliques to be empty.
    Next, we iterate over all possible colorings of the nodes in~$\mathcal{C}$ using~$\ell+1$ colors~$0,1,2,\ldots,\ell$.
    Note that there are at most~${(\ell+1)^{4k} \in O((2k+1)^{4k})}$ such colorings.

    The idea behind the coloring is the following.
    All colors~$1,2,\ldots,\ell$ will correspond to the cliques~$S_1,S_2,\ldots,S_\ell$, that is, we try to find a solution where all (vertices in critical cliques corresponding to) nodes of the same color (except for color~$0$) belong to the same clique in the solution.
    The color~$0$ indicates that the node will belong to multiple cliques in the solution, that is, all vertices in the respective critical clique will be split.
    Since each such split operation reduces~$k$ by one, we can reject any coloring in which the number of vertices in critical cliques corresponding to nodes with color~$0$ is more than~$k$.
    In particular, we can reject any coloring in which more than~$k$ nodes have color~$0$.

    Next, we guess two indices~$i \in [k], j \in [\ell]$ and assume that the~$i$\textsuperscript{th} node of color~$0$ belongs to~$S_j$ or we guess that all nodes of color~$0$ have been assigned to all cliques they belong to.
    Note that in each iteration, there are~$k\ell+1$ possibilities and since each guess reduces~$k$ by at least one, is the last guess corresponding to one of the nodes of color~$0$, or the last guess in general, we can make at most~$2k+1$ guesses.
    Hence, there are at most~$(k\ell+1)^{2k+1} = (2k^2+1)^{2k+1} \in O((2k+1)^{4k+1})$ such guesses.

    It remains to compute the best solution corresponding to each possible set of guesses.
    To this end, we first iterate over each pair of vertices and add an edge between them if this edge does not already exist and we guessed that there is a clique~$S_i$ which contains a copy of each of the two vertices.
    Moreover, we remove an existing edge between them if we guessed that the two vertices do not appear in a common clique.
    Finally, we perform all split operations.
    Therein, we iteratively split one vertex~$v$ into two vertices~$u_1$ and~$u_2$ where~$u_1$ will be the vertex in some clique~$S_i$ and~$u_2$ might be split further in the future.
    The vertex~$u_1$ is adjacent to all vertices that are guessed to belong to~$S_i$.
    The vertex~$u_2$ is adjacent to all vertices that~$u$ was adjacent to, except for vertices that are adjacent to~$u_1$ and not guessed to also belong to some other clique~$S_j$ which (some copy of)~$u_2$ belongs to.

    Since our algorithm basically performs an exhaustive search, it will find an optimal solution.
    It only remains to analyze the running time.
    We first compute the kernel in~$O(n+m)$ time.
    We then try~$O((2k+1)^{4k})$ possible colorings of~$\mathcal{C}$ and for each coloring~$O((2k+1)^{4k+1})$~guesses.
    Afterwards, we compute the solution in~$O(k^2)$~time as~$n \in O(k)$ by Reduction Rule~3.
    Thus, the overall running time is in~$O((2k+1)^{8k+1} \cdot k^2 + n+m) \subseteq O(2^{9k\log k}+n+m)$.
\ifarxiv\else\qed\fi
\end{proof}

We should note in passing that, while the constants in the running time of our algorithm can probably be improved, a completely new approach is required if one wants a single-exponential-time algorithm.
This is due to the fact that the number of possible partitions of~$O(k)$ critical cliques into clusters grows super-exponentially (roughly as fast as~$k!$) even if no vertex-splitting operations are allowed.
 
\section{Conclusion}
\label{sec:conclusion}

By allowing a vertex to split into two vertices, we extend the notion of \textsc{Cluster Editing} in an attempt to better model clustering problems where a data element may have roles in more than one cluster.
On the one hand, we show that this new problem, which we call \CEVS, is \NP-complete and, assuming the ETH, there are no $2^{o(n+m)}$-time or~$2^{o(k)} \cdot \poly(n)$-time algorithms for it.
On the other hand, we give a~$6k$-vertex kernel and an~$O(2^{9k \log k} + n + m)$-time algorithm.
This leaves the following gaps.

\begin{open}
  Does there exist a $2^{O(k)} \cdot \poly(n)$-time algorithm for \CEVS?
\end{open}

\begin{open}
  Does there exist a linear-size kernel, that is, a kernel in which the number of vertices plus the number of edges is in~$O(k)$?
\end{open}

\noindent
However, even resolving these questions should only be seen as a starting point for a much larger undertaking.
While we do understand the parameterized complexity of \CEVS{} with respect to~$k$ reasonably well, there are still a lot of open questions regarding structural parameters of the input graph.
Future work may also consider a bound on the number of allowed edge edits incident to each vertex as used by \citet{komusiewicz2012clusterediting} and~\citet{abu2017complexity}.
Moreover, one might also study the approximability of \CEVS{} as the trivial constant-factor approximation of \textsc{Cluster Editing} does not carry over if we allow vertex splitting.
In case \CEVS{} turns out to be hard to approximate, one might then continue with studying FPT-approximation (schemes) and approximation kernels (also known as lossy kernels).


The vertex splitting operation may also be applicable to other classes of target graphs, including bipartite graphs, chordal-graphs, comparability graphs, perfect graphs, or disjoint unions of graphs like complete bipartite graphs (bi-clusters),~$s$-cliques, $s$-clubs, $s$-plexes, $k$-cores, or~$\gamma$-quasi-cliques.
We note that the results in \cref{sec:editseq} are directly applicable to these settings as well.
Especially \textsc{Bicluster Editing} has received significant attention recently~\cite{drange2015fastbiclustering,guo2008improvedalgorithms,tsur2021fasterparameterized,xiao2022simpleimproved}.
To the best of our knowledge, nothing is known about \textsc{Bicluster Editing with Vertex Splitting}.
We should also note that there are two natural versions in the bipartite case and both of them seem worth studying.
The two versions differ in whether or not one requires that all copies of a split vertex lie on the same side of a bipartition in a solution.
On the one hand, the additional requirement makes sense if the data is inherently bipartite.
This happens for example if each vertex represents either a researcher or a paper and an edge represents an authorship.
On the other hand, if edges reflect something like a seller-buyer relationship, then it is plausible that a person both sells and buys.

Finally, we believe that it also makes sense to study a variant of the vertex-splitting operation where the neighborhood of the two newly introduced vertices are a partition of the neighborhood of the split vertex rather than a covering.
That is, when we split a vertex~$v$ into~$v_1$ and~$v_2$, instead of allowing a neighbor~$w$ of~$v$ to be adjacent to both~$v_1$ and~$v_2$, we require that~$w$ must be adjacent to \emph{exactly} one of the two. 
This operation is called \emph{exclusive vertex splitting} in the literature, and can be seen to be closely related to the \textsc{Clique Partitioning} problem.

\section*{Acknowledgments}
\noindent
Matthias Bentert is supported by the European Research Council (ERC) under the European Union’s Horizon 2020 research and innovation programme (grant agreement No. 819416).
Pål Grønås Drange was partially supported by the Research Council of Norway, grant Parameterized Complexity for Practical Computing (PCPC) (NFR, no.\ 274526).
Serge Gaspers is the recipient of an Australian Research Council (ARC) Future Fellowship (FT140100048) and acknowledges support under the ARC's Discovery Projects funding scheme (DP150101134).
Alexis Shaw is the recipient of an Australian Government Research Training Program Scholarship.

\bibliographystyle{abbrvnat}
\bibliography{references}

\end{document}

%% file: figwheel.tex
\begin{figure}[t]
  \centering
  \begin{subfigure}[b]{0.35\textwidth}
    \centering
    \resizebox{\linewidth}{!}{
      \begin{tikzpicture}[scale=0.8,every node/.style={circle,draw,scale=0.8}]
        \graph[clockwise, radius=3cm, n=6,empty nodes]
        {
          v1,v2,v3,v4,v5,v6
        };
        \coordinate (CENTER1) at ($(v1)!0.7!(v3)$); 
        \coordinate (CENTER2) at ($(v4)!0.7!(v6)$); 
        \node[draw] (c1) [at=(CENTER1)] {};
        \node[draw] (c2) [at=(CENTER2)] {};
        \draw (c1) -- (c2);
        \draw (v1) -- (v2) -- (v3) -- (v4) -- (v5) -- (v6) -- (v1); 
        \draw (c1) -- (v1);
        \draw (c1) -- (v2);
        \draw (c1) -- (v3);
        \draw (c1) -- (v4);
        \draw (c1) -- (v5);
        \draw (c1) -- (v6);
        \draw (c2) -- (v1);
        \draw (c2) -- (v2);
        \draw (c2) -- (v3);
        \draw (c2) -- (v4);
        \draw (c2) -- (v5);
        \draw (c2) -- (v6);
      \end{tikzpicture}
    }
    \caption{The graph~$W_{6}$ (variable gadget).}
  \end{subfigure}
  \hspace{2em}
    \begin{subfigure}[b]{0.35\textwidth}
    \centering
    \resizebox{\linewidth}{!}{
      \begin{tikzpicture}[scale=0.8,every node/.style={circle,draw,scale=0.8}]
        \graph[clockwise, radius=3cm, n=6, empty nodes]
        {
          v1,v2,v3,v4,v5,v6
        };
%

        \coordinate (CENTER1) at ($(v1)!0.7!(v3)$);
        \coordinate (CENTER3) at ($(v2)!0.55!(v4)$);
        \coordinate (CENTER2) at ($(v4)!0.7!(v6)$);
        \coordinate (CENTER4) at ($(v5)!0.55!(v1)$);
        \node[draw] (c1) [at=(CENTER1)] {};
        \node[draw,added] (c3) [at=(CENTER3)] {};
		\node[draw] (c2) [at=(CENTER2)] {};
        \node[draw,added] (c4) [at=(CENTER4)] {};

        \draw (c1) -- (c4);
        \draw (c3) -- (c2);

        \draw (v1) -- (v2) -- (v3);
        \draw (v4) -- (v5) -- (v6);
        
        \draw[deleted] (v6) -- (v1);
        \draw[deleted] (v3) -- (v4);
        
        \draw[added,bend left=15] (v1) to (v3);
        \draw[added,bend left=15] (v4) to (v6);
        
        \draw (c1) -- (v1);
        \draw (c1) -- (v2);
        \draw (c1) -- (v3);

        \draw (c3) -- (v4);
        \draw (c3) -- (v5);
        \draw (c3) -- (v6);

        \draw (c4) -- (v1);
        \draw (c4) -- (v2);
        \draw (c4) -- (v3);

        \draw (c2) -- (v4);
        \draw (c2) -- (v5);
        \draw (c2) -- (v6);
      \end{tikzpicture}
    }
    \caption{One of the three ways of transforming~$W_6$ into two~$K_5$'s using six operations.}
  \end{subfigure}
\caption{The graph~$W_{6t}$ requires~$8t - 2$ edits and any solution with exactly~$8t-2$ edits results in a disjoint union of~$K_5$s.}
    \label{fig:wheel}
\end{figure}


%% file: figcrown.tex
\begin{figure}[t]
  \centering
  \begin{subfigure}[b]{0.28\textwidth}
    \centering
    \resizebox{\linewidth}{!}{
    \begin{tikzpicture}[scale=0.8,every node/.style={circle,draw,scale=0.8},node distance=2cm]
      \node (a) {};
      \node (b) [right of=a] {};
      \coordinate (CENTER) at ($(a)!0.5!(b)$);
      \node (y) [above of=CENTER] {$y$};
      \node (x) [left of=y] {$x$};
      \node (z) [right of=y] {$z$};
      \draw (a) -- (b) -- (x) -- (a);
      \draw (a) -- (y) -- (b);
      \draw (a) -- (z) -- (b);
    \end{tikzpicture}
    }
    \caption{The crown graph}
  \end{subfigure}
  \hspace{3em}
  \begin{subfigure}[b]{0.28\textwidth}
    \centering
    \resizebox{\linewidth}{!}{
      \begin{tikzpicture}[scale=0.8,every node/.style={circle,draw,scale=0.8},node distance=2cm]
        \node (a) {};
        \node (b) [right of=a] {};
        \coordinate (CENTER) at ($(a)!0.5!(b)$);
        \node (y) [above of=CENTER] {$y$};
        \node (x) [left of=y] {$x$};
        \node (z) [right of=y] {$z$};
        \draw (a) -- (b) -- (x) -- (a);
        \draw (a) -- (y) -- (b);
        \draw[added] (x) -- (y);
        \draw[deleted] (a) -- (z) -- (b);
      \end{tikzpicture}
    }
    \caption{Good solution: One added edge and two deleted edges}
    \label{fig:crown-good}
  \end{subfigure}

  \begin{subfigure}[b]{0.28\textwidth}
    \centering
    \resizebox{\linewidth}{!}{
      \begin{tikzpicture}[scale=0.8,every node/.style={circle,draw,scale=0.8},node distance=2cm]
        \node (a) {};
        \node (b) [right of=a] {};
        \coordinate (CENTER) at ($(a)!0.5!(b)$);
        \node (y) [above of=CENTER] {$y$};
        \node (x) [left of=y] {$x$};
        \node (z) [right of=y] {$z$};
        \draw (a) -- (b) -- (x) -- (a);
        \draw (a) -- (y) -- (b);
        \draw (a) -- (z) -- (b);
        \draw[added] (x) -- (y) -- (z);
        \draw[added] (x) to[bend left] (z);
      \end{tikzpicture}
    }
    \caption{Bad solution 1: three added edges}
    \label{fig:crown-bad-1}
  \end{subfigure}
  \hspace{2em}
  \begin{subfigure}[b]{0.28\textwidth}
    \centering
    \resizebox{\linewidth}{!}{
      \begin{tikzpicture}[scale=0.8,every node/.style={circle,draw,scale=0.8},node distance=2cm]
        \node (a) {};
        \node (b) [right of=a] {};
        \node (a1) [added, right=0.1cm of b]{};
        \node (b1) [added, right=0.1cm of a1]{};
        \coordinate (CENTER) at ($(a)!0.5!(b)$);
        \node (y) [above of=CENTER] {$y$};
        \node (x) [left of=y] {$x$};
        \node (z) [right of=y] {$z$};
        \draw (a) -- (b) -- (x) -- (a);
        \draw (a) -- (y) -- (b);
        \draw[added] (x) -- (y);
        \draw (z) -- (a1) -- (b1) -- (z);
      \end{tikzpicture}
    }
    \caption{Bad solution 2: two splits and one added edge}
    \label{fig:crown-bad-2}
  \end{subfigure}
  \hspace{2em}
  \begin{subfigure}[b]{0.28\textwidth}
    \centering
    \resizebox{\linewidth}{!}{
      \begin{tikzpicture}[scale=0.8,every node/.style={circle,draw,scale=0.8},node distance=2cm]
        \node (a) {};
        \node (b) [right of=a] {};
        \node (b1) [added, right=0.5cm of b]{};
        \coordinate (CENTER) at ($(a)!0.5!(b)$);
        \node (y) [above of=CENTER] {$y$};
        \node (x) [left of=y] {$x$};
        \node (z) [right of=y] {$z$};
        \draw (a) -- (b) -- (x) -- (a);
        \draw (a) -- (y) -- (b);
        \draw[added] (x) -- (y);
        \draw (z) -- (b1);
        \draw[deleted] (a) -- (z);
      \end{tikzpicture}
    }
    \caption{Bad solution 3: one split, one deleted edge, and one added edge}
    \label{fig:crown-bad-3}
  \end{subfigure}
  \caption{The crown graph with its four solutions of size 3.  The
    \emph{good solution} is the only solution with three operations that
    creates at least one isolated vertex.}
    \label{fig:crown}
\end{figure}


%% file: main.bbl
\begin{thebibliography}{31}
\providecommand{\natexlab}[1]{#1}
\providecommand{\url}[1]{\texttt{#1}}
\expandafter\ifx\csname urlstyle\endcsname\relax
  \providecommand{\doi}[1]{doi: #1}\else
  \providecommand{\doi}{doi: \begingroup \urlstyle{rm}\Url}\fi

\bibitem[Abu{-}Khzam(2017)]{abu2017complexity}
F.~N. Abu{-}Khzam.
\newblock On the complexity of multi-parameterized cluster editing.
\newblock \emph{Journal of Discrete Algorithms}, 45:\penalty0 26--34, 2017.

\bibitem[Abu-Khzam et~al.(2005)Abu-Khzam, Baldwin, Langston, and
  Samatova]{abu2005}
F.~N. Abu-Khzam, N.~E. Baldwin, M.~A. Langston, and N.~F. Samatova.
\newblock On the relative efficiency of maximal clique enumeration algorithms,
  with applications to high-throughput computational biology.
\newblock In \emph{Proceedings of the 2005 International Conference on Research
  Trends in Science and Technology}, pages 1--10, 2005.

\bibitem[Abu-Khzam et~al.(2018)Abu-Khzam, Egan, Gaspers, Shaw, and
  Shaw]{abu2018cluster}
F.~N. Abu-Khzam, J.~Egan, S.~Gaspers, A.~Shaw, and P.~Shaw.
\newblock Cluster editing with vertex splitting.
\newblock In \emph{International Symposium on Combinatorial Optimization},
  pages 1--13. Springer, 2018.

\bibitem[Arrighi et~al.(2023)Arrighi, Bentert, Drange, Sullivan, and
  Wolf]{AMGSW2023}
E.~Arrighi, M.~Bentert, P.~G. Drange, B.~D. Sullivan, and P.~Wolf.
\newblock Cluster editing with overlapping communities.
\newblock In \emph{Proceedings of the 18th {International} {Symposium} on
  {Parameterized} and {Exact} {Computation} ({IPEC~'23})}. Schloss Dagstuhl —
  Leibniz-Zentrum für Informatik, 2023.
\newblock To appear.

\bibitem[B{\"o}cker(2012)]{Bocker12}
S.~B{\"o}cker.
\newblock A golden ratio parameterized algorithm for cluster editing.
\newblock \emph{Journal of Discrete Algorithms}, 16:\penalty0 79--89, 2012.

\bibitem[Cai(1996)]{Cai96}
L.~Cai.
\newblock Fixed-parameter tractability of graph modification problems for
  hereditary properties.
\newblock \emph{Information Processing Letters}, 58\penalty0 (4):\penalty0
  171--176, 1996.

\bibitem[Chen and Meng(2012)]{ChenMeng}
J.~Chen and J.~Meng.
\newblock A 2$k$ kernel for the cluster editing problem.
\newblock \emph{Journal of Computer and System Sciences}, 78\penalty0
  (1):\penalty0 211--220, 2012.

\bibitem[Chen et~al.(2006)Chen, Huang, Kanj, and Xia]{Chen2006}
J.~Chen, X.~Huang, I.~A. Kanj, and G.~Xia.
\newblock Strong computational lower bounds via parameterized complexity.
\newblock \emph{Journal of Computer and System Sciences}, 72\penalty0
  (8):\penalty0 1346 -- 1367, 2006.

\bibitem[Cygan et~al.(2015)Cygan, Fomin, Kowalik, Lokshtanov, Marx, Pilipczuk,
  Pilipczuk, and Saurabh]{CFKLMPPS15}
M.~Cygan, F.~V. Fomin, L.~Kowalik, D.~Lokshtanov, D.~Marx, M.~Pilipczuk,
  M.~Pilipczuk, and S.~Saurabh.
\newblock \emph{Parameterized Algorithms}.
\newblock Springer, 2015.

\bibitem[D{'}Addario et~al.(2014)D{'}Addario, Kopczynski, Baumbach, and
  Rahmann]{DAddario2014}
M.~D{'}Addario, D.~Kopczynski, J.~Baumbach, and S.~Rahmann.
\newblock A modular computational framework for automated peak extraction from
  ion mobility spectra.
\newblock \emph{BMC Bioinformatics}, 15\penalty0 (1):\penalty0 25, 2014.

\bibitem[Dehne et~al.(2006)Dehne, Langston, Luo, Pitre, Shaw, and
  Zhang]{dehne2006cluster}
F.~Dehne, M.~A. Langston, X.~Luo, S.~Pitre, P.~Shaw, and Y.~Zhang.
\newblock The cluster editing problem: Implementations and experiments.
\newblock In \emph{Proceedings of the 2nd International Workshop on
  Parameterized and Exact Computation (IWPEC~'06)}, pages 13--24. Springer
  Berlin Heidelberg, 2006.

\bibitem[Diestel(2005)]{diestel2005graphtheory}
R.~Diestel.
\newblock \emph{Graph Theory}.
\newblock Springer, 2005.

\bibitem[Drange et~al.(2015)Drange, Reidl, Villaamil, and
  Sikdar]{drange2015fastbiclustering}
P.~G. Drange, F.~Reidl, F.~S. Villaamil, and S.~Sikdar.
\newblock Fast biclustering by dual parameterization.
\newblock In \emph{Proceedings of the 10th {International} {Symposium} on
  {Parameterized} and {Exact} {Computation} ({IPEC~'15})}, pages 402--413.
  Schloss Dagstuhl — Leibniz-Zentrum für Informatik, 2015.

\bibitem[Fadiel et~al.(2006)Fadiel, Langston, Peng, Perkins, Taylor, Tuncalp,
  Vitello, Pevsner, and Naftolin]{fadiel2006computational}
A.~Fadiel, M.~A. Langston, X.~Peng, A.~D. Perkins, H.~S. Taylor, O.~Tuncalp,
  D.~Vitello, P.~H. Pevsner, and F.~Naftolin.
\newblock Computational analysis of mass spectrometry data using novel
  combinatorial methods.
\newblock In \emph{Proceedings of the 2006 {IEEE/ACS} International Conference
  on Computer Systems and Applications {(AICCSA}~'06)}, pages 266--273. {IEEE}
  Computer Society, 2006.

\bibitem[Fellows et~al.(2007)Fellows, Langston, Rosamond, and
  Shaw]{fellows2007efficient}
M.~Fellows, M.~Langston, F.~Rosamond, and P.~Shaw.
\newblock Efficient parameterized preprocessing for cluster editing.
\newblock In \emph{Proceedings of the 16th International Symposium on
  Fundamentals of Computation Theory {(FCT~'07)}}, pages 312--321. Springer,
  2007.

\bibitem[Firbas et~al.(2023)Firbas, Dobler, Holzer, Schafellner, Sorge,
  Villedieu, and Wi{\ss}mann]{FDH+23}
A.~Firbas, A.~Dobler, F.~Holzer, J.~Schafellner, M.~Sorge, A.~Villedieu, and
  M.~Wi{\ss}mann.
\newblock The complexity of cluster vertex splitting and company.
\newblock \emph{CoRR}, abs/2309.00504, 2023.

\bibitem[Flum and Grohe(2006)]{flum2006parameterizedcomplexity}
J.~Flum and M.~Grohe.
\newblock \emph{Parameterized {Complexity} {Theory}}.
\newblock Springer, 2006.

\bibitem[Gramm et~al.(2005)Gramm, Guo, H{\"u}ffner, and Niedermeier]{Gramm05}
J.~Gramm, J.~Guo, F.~H{\"u}ffner, and R.~Niedermeier.
\newblock Graph-modeled data clustering: Exact algorithms for clique
  generation.
\newblock \emph{Theory of Computing Systems}, 38\penalty0 (4):\penalty0
  373--392, 2005.

\bibitem[Guo(2009)]{guo2009more}
J.~Guo.
\newblock A more effective linear kernelization for cluster editing.
\newblock \emph{Theoretical Computer Science}, 410\penalty0 (8-10):\penalty0
  718 -- 726, 2009.

\bibitem[Guo et~al.(2008)Guo, Hüffner, Komusiewicz, and
  Zhang]{guo2008improvedalgorithms}
J.~Guo, F.~Hüffner, C.~Komusiewicz, and Y.~Zhang.
\newblock Improved algorithms for bicluster editing.
\newblock In \emph{Proceedings of the 5th International Conference on {Theory}
  and {Applications} of {Models} of {Computation} ({TAMC~'08})}, pages
  445--456. Springer, 2008.

\bibitem[H{\"u}ffner et~al.(2010)H{\"u}ffner, Komusiewicz, Moser, and
  Niedermeier]{Huffner2010}
F.~H{\"u}ffner, C.~Komusiewicz, H.~Moser, and R.~Niedermeier.
\newblock Fixed-parameter algorithms for cluster vertex deletion.
\newblock \emph{Theory of Computing Systems}, 47\penalty0 (1):\penalty0
  196--217, 2010.

\bibitem[Impagliazzo et~al.(2001)Impagliazzo, Paturi, and
  Zane]{impagliazzo2001whichproblems}
R.~Impagliazzo, R.~Paturi, and F.~Zane.
\newblock Which problems have strongly exponential complexity?
\newblock \emph{Journal of Computer and System Sciences}, 63\penalty0
  (4):\penalty0 512--530, 2001.

\bibitem[Komusiewicz and Uhlmann(2012)]{komusiewicz2012clusterediting}
C.~Komusiewicz and J.~Uhlmann.
\newblock Cluster editing with locally bounded modifications.
\newblock \emph{Discrete Applied Mathematics}, 160\penalty0 (15):\penalty0
  2259--2270, 2012.

\bibitem[K{\u r}iv{\'a}nek and Mor{\'a}vek(1986)]{KM86}
M.~K{\u r}iv{\'a}nek and J.~Mor{\'a}vek.
\newblock {NP}-hard problems in hierarchical-tree clustering.
\newblock \emph{Acta Informatica}, 23\penalty0 (3):\penalty0 311--323, 1986.

\bibitem[Lin et~al.(2000)Lin, Jiang, and Kearney]{lin2000phylogenetic}
G.-H. Lin, T.~Jiang, and P.~E. Kearney.
\newblock Phylogenetic $k$-root and {S}teiner $k$-root.
\newblock In \emph{Proceedings of the 11th {International} {Symposium} on
  {Algorithms} and {Computation} ({ISAAC}~'00)}, pages 539--551. Springer,
  2000.

\bibitem[Niedermeier(2006)]{Niedermeier06}
R.~Niedermeier.
\newblock \emph{An Invitation to Fixed-Parameter Algorithms}.
\newblock Oxford University Press, 2006.

\bibitem[Radovanovi{\'c} et~al.(2010)Radovanovi{\'c}, Nanopoulos, and
  Ivanovi{\'c}]{radovanovic2010hubs}
M.~Radovanovi{\'c}, A.~Nanopoulos, and M.~Ivanovi{\'c}.
\newblock Hubs in space: Popular nearest neighbors in high-dimensional data.
\newblock \emph{Journal of Machine Learning Research}, 11:\penalty0 2487--2531,
  2010.

\bibitem[Tomasev et~al.(2014)Tomasev, Radovanovic, Mladenic, and
  Ivanovic]{tomavsev2011role}
N.~Tomasev, M.~Radovanovic, D.~Mladenic, and M.~Ivanovic.
\newblock The role of hubness in clustering high-dimensional data.
\newblock \emph{{IEEE} Transactions on Knowledge and Data Engineering},
  26\penalty0 (3):\penalty0 739--751, 2014.

\bibitem[Tovey(1984)]{tovey1984asimplified}
C.~A. Tovey.
\newblock A simplified {NP}-complete satisfiability problem.
\newblock \emph{Discrete Applied Mathematics}, 8\penalty0 (1):\penalty0 85--89,
  1984.

\bibitem[Tsur(2021)]{tsur2021fasterparameterized}
D.~Tsur.
\newblock Faster parameterized algorithm for bicluster editing.
\newblock \emph{Information Processing Letters}, 168, 2021.

\bibitem[Xiao and Kou(2022)]{xiao2022simpleimproved}
M.~Xiao and S.~Kou.
\newblock A simple and improved parameterized algorithm for bicluster editing.
\newblock \emph{Information Processing Letters}, 2022.

\end{thebibliography}
